\documentclass[a4paper]{jpconf}
\usepackage{geometry}
\usepackage{graphicx}
\begin{document}
\title{A predictive standard model for heavy electron systems}

\author{Yi-feng Yang$^1$, N J Curro$^2$, Z Fisk$^3$, D Pines$^2$, J D Thompson$^1$}

\address{$^1$ Los Alamos National Laboratory, Los Alamos, NM 87545, USA}
\address{$^2$ Department of Physics, University of California, Davis, CA 95616, USA}
\address{$^3$ Department of Physics and Astronomy, University of California, Irvine, CA 92697, USA}

\ead{yifengyyf@gmail.com}

\begin{abstract}
We propose a predictive standard model for heavy electron systems based on a detailed phenomenological two-fluid description of existing experimental data. It leads to a new phase diagram that replaces the Doniach picture, describes the emergent anomalous scaling behavior of the heavy electron (Kondo) liquid measured below the lattice coherence temperature, $T^*$, seen by many different experimental probes, that marks the onset of collective hybridization, and enables one to obtain important information on quantum criticality and the superconducting/antiferromagnetic states at low temperatures. Because $T^*$ is $\sim J^2\rho/2$, the nearest neighbor RKKY interaction, a knowledge of the single-ion Kondo coupling, $J$, to  the background conduction electron density of states, $\rho$, makes it possible to predict Kondo liquid behavior, and to estimate its  maximum superconducting transition temperature in both existing and  newly discovered heavy electron families. 
\end{abstract}

\section{Introduction}

Although we have yet to obtain a microscopic solution of the Kondo lattice problem that serves as a model for heavy electron materials, thanks to a recent phenomenological analysis \cite{Yang2008a,Yang2008b,Yang2009a,Yang2009b} of decades of experimental efforts on heavy electron behavior, we now have an experiment-based model for their behavior that explains existing experiments and has predictive power in the exploration of new families of heavy electron materials. Our model provides a unified explanation of the emergence of the heavy electron Kondo liquid that is responsible for the anomalies seen in the Knight shift, magneto-transport, and other experiments, leads to a new phase diagram that replaces the Doniach picture, and has the potential for becoming the standard model of heavy electron physics. It contains several basic aspects: 

\begin{itemize}
\item{There exists a unique temperature scale $T^*$ for each heavy electron material; it marks the emergence of a new state of matter, a coherent heavy electron (Kondo) liquid that is  produced by the collective hybridization of the local moments with the conduction electron sea, and exhibits universal behavior that scales with $T^*$\cite{Yang2008a}.}

\item{$T^*$ is the nearest neighbor RKKY (Ruderman-Kittel-Kasuya-Yosida) local moment interaction; for the local moments it separates their high-$T$ weakly interacting phase from their low-$T$ entangled phase. It is given by $\sim\,$$0.45J^2\rho$, where $J$ is the single-ion Kondo coupling to the background conduction electron density of states $\rho$ \cite{Yang2008b}.}

\item{The entangled phase below $T^*$ is described by a two-fluid model \cite{Yang2008a,Nakatsuji2004,Curro2004} in which an order parameter, $f(T/T^*)$, characterizes the reduction of the overall system entropy brought about by the transfer of spectral weight from the f-electron local moments to the itinerant  heavy electron Kondo liquid formed out of the conduction electron sea.}

\item{The quasiparticles of the Kondo liquid exhibit coherent non-Landau Fermi liquid behavior; their  spin susceptibility and specific heat vary as $\ln(T^*/T)$ down to a crossover temperature $T_0$.}

\item{Their behavior below $T_0$ can lead to unconventional superconductivity produced by quantum critical fluctuations \cite{Yang2009b}, or an antiferromagntic ordering of the local moments in which Kondo liquid relocalization occurs  as a precursor \cite{Yang2010}.}
\end{itemize}

\begin{figure}[h]
\begin{minipage}{18pc}
\includegraphics[width=18pc]{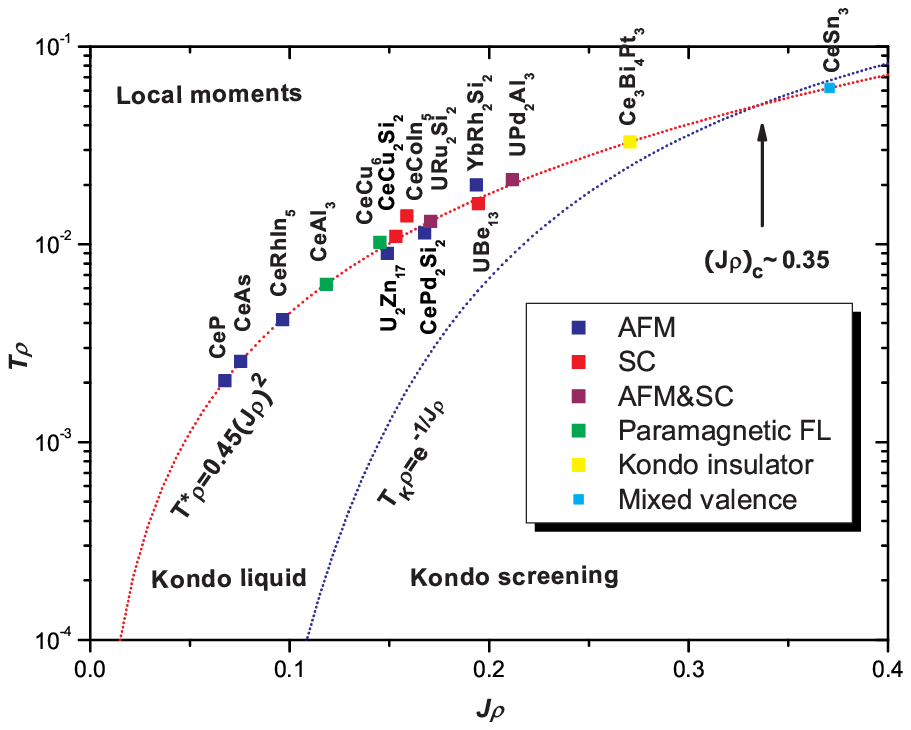}
\caption{\label{fig1}General phase diagram. $J$ is the local Kondo coupling and $\rho$ is the conduction electron density of states. $T^*$ marks the onset of coherence, spin correlations and anomalies.}
\end{minipage}\hspace{2pc}%
\begin{minipage}{18pc}
\includegraphics[width=18pc]{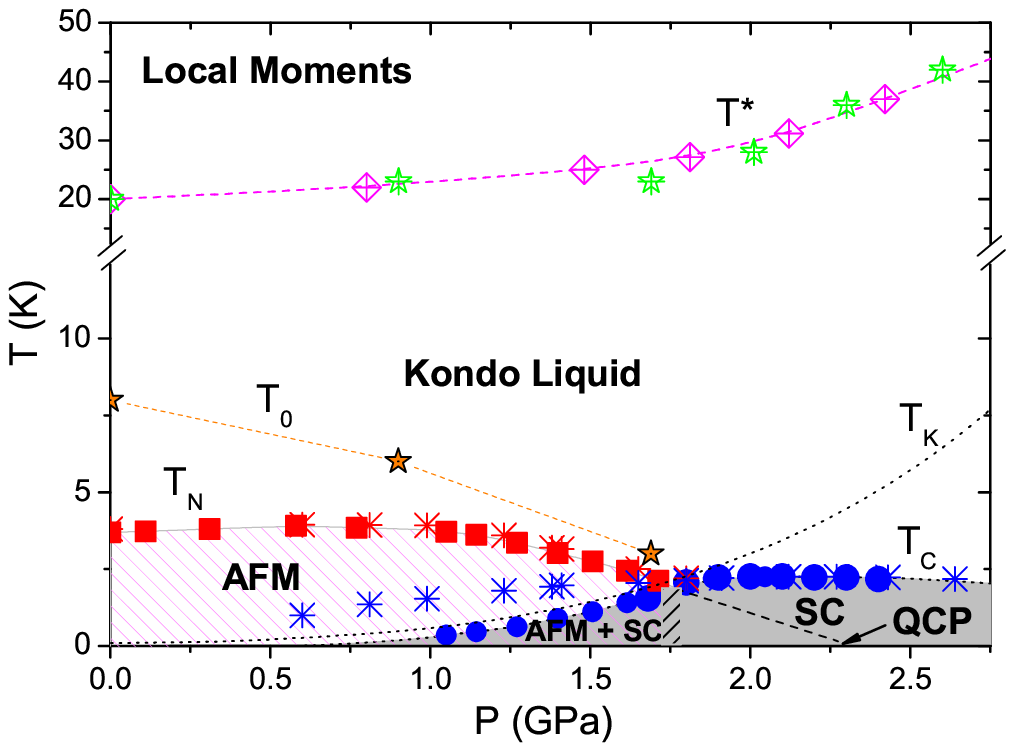}
\caption{\label{fig2}New phase diagram of CeRhIn$_5$ \cite{Park2009}. $T^*$ is from resistivity peak and Hall anomaly. $T_0$ is from Hall anomaly \cite{Nakajima2007}. $T_K$ is estimated from Ce$_{1-x}$La$_x$RhIn$_5$.}
\end{minipage} 
\end{figure}

The position of a number of heavy electron materials on our proposed new phase diagram is given in Fig.~\ref{fig1}, where we see that $J\rho\sim\,$0.15 roughly separates the  antiferromagnets and superconductors. In striking contrast to the conventional Doniach diagram which postulates  a competition between collective behavior produced by  RKKY interaction  and single-ion behavior characterized by the single-ion Kondo temperature, $T_K$, the latter quantity is  found to be much smaller than $T^*$ for almost all the materials considered there. In Fig.~\ref{fig2} we plot the variation of $T^*$, $T_0$, $T_K$, and the ordering temperatures, $T_N$ and $T_c$ as a function of pressure for the heavy electron material CeRhIn$_5$. There one can see how  the competition between localization (antiferromagnetic order)  and itinerancy (superconductivity) for Kondo liquid quasiparticles plays out as $J\rho$ increases.

\section{The physical origin and determination of the fundamental temperature scale $T^*$}

The Kondo coupling $J$ between a localized f-moment and conduction electron spins gives rise to two competing characteristic temperature scales in a Kondo lattice: 
\begin{eqnarray}
k_BT_K&=&\rho^{-1}e^{-1/J\rho},\\
k_BT_{RKKY}&=&cJ^2\rho,
\label{RKKY}
\end{eqnarray}
where $\rho$, the density of states of conduction electrons, can be determined by the Sommerfeld coefficient $\gamma$ of the corresponding nonmagnetic compound (e.g., LaCoIn$_5$ for CeCoIn$_5$), $\gamma=\pi^2k_B^2\rho/3$.  After a correction to account for change of chemical pressure upon doping, the value of $J$ for the Kondo lattice material can be determined by measurements of $T_K$  in the diluted compound (e.g., Ce$_{1-x}$La$_x$CoIn$_5$), where it is given by  $\pi^2k_B^2/3\gamma\ln(\pi^2k_B/3\gamma T_K)$.  An analysis \cite{Yang2008b} of dozens of heavy electron materials and the pressure data of CeRhIn$_5$ showed that in all cases of interest, $T_K$ was quite small compared to $T^*$, so that it was natural to inquire whether $T^*$ might be given by the RKKY interaction, Eq.~(\ref{RKKY}), with a near-universal value of $c$. As may be seen in Fig.~\ref{fig1}, this is indeed the case, with a best fit analysis yielding $c\sim\,$0.45, so that  
\begin{equation}
k_BT^*=0.45J^2\rho.
\end{equation}

\section{The emergent heavy electron liquid}

Although $T^*$ marks the onset of coherent collective hybridization, the process is not instantaneous; the heavy electron Kondo liquid emerges over a broad temperature range from $T^*$ to a much lower cut-off temperature, $T_0$, and throughout  this temperature region, one therefore expects  two components: the emergent Kondo non-Landau  Fermi liquid composed of  conduction electrons made heavy by their collective hybridization with the local moments and a  local moment spin liquid formed by the  substantial fraction of local moments that have not hybridized.

One is thus led to postulate a two-fluid model \cite{Nakatsuji2004} with an order parameter $f(T/T^*)$ that denotes  the fraction of the f-moments that participate in the collective hybridization;  both $f(T/T^*)$ and the scaling properties of the Kondo liquid can be determined directly from experiments displaying the various heavy electron anomalies. Consider, for example, Knight shift measurements \cite{Curro2004}. Since the f-moments are fully localized above $T^*$, the Knight shift of a given probe nucleus measures directly  their susceptibility. Below $T^*$, it probes both the local moment and Kondo liquid contributions to the total magnetic susceptibility, $\chi$, But since it couples differently to each, it will no longer be proportional to the total susceptibility, hence the appearance of the  anomaly seen beginning at $T^*$. As shown in \cite{Yang2008a}, the two-fluid model expressions are thus:
\begin{eqnarray}
\chi&=&f(T)\chi_h + [1-f(T)]\chi_l,\\
K&=&K_0+Af(T)\chi_h + B[1-f(T)]\chi_l,
\end{eqnarray}
where $\chi_h$ and $\chi_l$ are the intrinsic susceptibility of the heavy electrons and the unhybridized local moments, respectively. $A$ and $B$ are their hyperfine couplings. Since the Knight shift anomaly below $T^*$ is given by,
\begin{equation}
K_a=K-K_0-B\chi=(A-B)f(T)\chi_h,
\end{equation}
we see that it acts as a direct probe of the emergent Kondo liquid susceptibility. A careful analysis of the Knight shift and other experiments on CeCoIn$_5$  shows that
\begin{equation}
f(T)=f_0\left(1-\frac{T}{T^*}\right)^{3/2},
\end{equation}
while if one requires that the intrinsic entropy of the Kondo liquid approaches the free moment entropy $R\ln$2 at $T^*$,
\begin{equation}
\chi_h=R_W\frac{R\ln2}{2T^*}\left(1+\ln\frac{T^*}{T}\right),
\label{chih}
\end{equation}
where $R_W$ is the Wilson ratio.  Knight shift measurements on other heavy electron materials and fits to the anomalous behavior seen in magneto-transport, point contact spectroscopy, and Raman spectroscopy show that these expressions are universal \cite{Yang2008a}. A comparison to recent state-of-the-art numerical DMFT calculations \cite{Shim2007} shows good agreement with the Kondo liquid quasi-particle density of states determined from Eq.~(\ref{chih}), and demonstrates that that these contain the feedback needed to achieve collective hybridization.

\section{Superconducting condensation and antiferromagnetic relocalization}

The low temperature cut-off at $T_0$ in the measured scaling behavior of the Kondo liquid marks the onset of new physical processes that take precedence over its growth due to collective hybridization. The two-fluid model continues to apply in this region, and provides important information about both the unconventional superconducting states and the onset of antiferromagnetism observed in heavy electron materials. Knight shift measurements are especially informative, as these provide direct information on changes in the Kondo liquid as the material becomes superconducting or antiferromagnetic. An analysis of CeCoIn$_5$ confirms the universal behavior of the Kondo liquid \cite{Yang2009b}: it follows two-fluid prediction above $T_0$ ($\sim T_c$) and, upon becoming superconducting, follows  the  BCS d-wave prediction below $T_c$, independent of the probe nuclei. Important information on the pairing mechanism for the measured d-wave superconductivity can be obtained from the NQR spin-lattice relaxation rate $T_1^{-1}$.
by using the two-fluid formula 
\begin{equation}
\frac{1}{T_1}=\frac{f(T)}{T_{1h}}+\frac{1-f(T)}{T_{1l}},
\end{equation}
where the local moment $T_{1l}^{-1}$ is given by the high temperature fit. 
A simple two-fluid analysis reveals quantum critical fluctuations in the Kondo liquid, and yields \cite{Yang2009b}
\begin{equation}
T_{1h}T\propto (T+T_{QCP}),
\end{equation}
where $T_{QCP}$ is the distance from the quantum critical point. 

A similar two-fluid analysis applied to heavy electron antiferromagnets CeRhIn$_5$ and CePt$_2$In$_7$ has recently led to the discovery that at a temperature $T_0$ well above the N\'eel temperature, the Kondo liquid begins to relocalize, with a corresponding decrease in $f(T)$, confirming the important role played by the Kondo liquid in bringing about the antiferromagnetic order of both it and the residual local moments present at $T_N$.

\section{Toward a standard predictive model of heavy electron materials}

By analogy to the idea of a standard model based on a well-defined microscopic theory, we have proposed here an experiment-based standard phenomenological model for heavy electron materials that derives from the successful systematic examination of existing experiments on a broad spectrum of heavy electron materials. Its predictions for heavy electron emergence at $T^*$ accompanied by anomalous scaling behavior can easily be checked for any newly-discovered heavy electron family, and have recently been confirmed by experiments on the newly-discovered 127 family.

Of particular interest are the correlations in our model between measured values of $J\rho$ and the appearance of antiferromagnetism and superconductivity at low temperatures. As we have noted, materials with $J\rho<0.15$ all appear to be antiferromagnets, but can easily be made into superconductors by applying pressure to increase $J\rho$, as Fig.~\ref{fig2} demonstrates vividly for the Rh-115 family. Moreover, since the superconducting transition temperature seems to scale with $T^*$, a strategy for finding a higher $T_c$ family of heavy electron materials might begin by searching for materials with large values of $T^*$, a strategy that proved successful in the discovery of the PuGa members of the 115 family. We note this can be achieved either by finding materials with values of $J\rho$ that significantly exceed 0.15, or, with $J\rho\sim0.15$, finding materials that  possess a comparatively low value of $\rho$, or equivalently a large conduction electron bandwidth, and a correspondingly high value of $J$. Moreover, to the extent that $T_c$ scales with $T^*$,  for a fixed value of $J\rho$, one would have $T_c$ scaling with $J$ or the conduction electron bandwidth $W\sim\rho^{-1}$. 

\ack
Y.Y. thanks H.-O. Lee and T. Park for experimental data of CeRhIn$_5$. Z.F. acknowledges support through NSF Grant No. NSF-DMR-0801253. Work at Los Alamos was performed under the auspices of the US Department of Energy through the LDRD program.

\end{document}